\documentclass[sigplan,screen, authorversion, nonacm]{acmart}
\AtBeginDocument{%
  \providecommand\BibTeX{{%
    \normalfont B\kern-0.5em{\scshape i\kern-0.25em b}\kern-0.8em\TeX}}}
\usepackage[frozencache,cachedir=.]{minted}
\setminted{fontsize=\scriptsize}
\usepackage{verbatim}
\usepackage{multirow}
\usepackage{tcolorbox}
\usepackage{adjustbox}
\usepackage{listings}
\DeclareCaptionStyle{ruled}{labelfont=normalfont,labelsep=colon,strut=off} 
\floatstyle{ruled}
\newfloat{listing}{tb}{lst}{}
\floatname{listing}{Listing}

\begin{document}

\title{Copilot Evaluation Harness: Evaluating LLM-Guided Software Programming}


\author{Anisha Agarwal}
\affiliation{%
\institution{Microsoft}
  \state{Redmond}
  \country{USA}
}
\author{Aaron Chan}
\affiliation{%
\institution{Microsoft}
  \state{Redmond}
  \country{USA}
}
\author{Shubham Chandel}
\affiliation{%
\institution{Microsoft}
  \state{Redmond}
  \country{USA}
}
\author{Jinu Jang}
\affiliation{%
\institution{Microsoft}
  \state{Redmond}
  \country{USA}
}

\author{Shaun Miller}
\affiliation{%
\institution{Microsoft}
  \state{Redmond}
  \country{USA}
}
\author{Roshanak Zilouchian Moghaddam}
\affiliation{%
\institution{Microsoft}
  \state{Redmond}
  \country{USA}
}

\author{Yevhen Mohylevskyy}
\affiliation{%
\institution{Microsoft}
  \state{Redmond}
  \country{USA}
}
\author{Neel Sundaresan}
\affiliation{%
\institution{Microsoft}
  \state{Redmond}
  \country{USA}
}
\author{Michele Tufano}
\affiliation{%
\institution{Microsoft}
  \state{Redmond}
  \country{USA}
}

\renewcommand{\shortauthors}{Agarwal and Chan, et al.}
\begin{abstract}
 The integration of Large Language Models (LLMs) into Development Environments (IDEs) has become a focal point in modern software development. LLMs such as OpenAI GPT-3.5/4 and Code Llama offer the potential to significantly augment developer productivity by serving as intelligent, chat-driven programming assistants. However, utilizing LLMs out of the box is unlikely to be optimal for any given scenario. 
 Rather, each system requires the LLM to be honed to its set of heuristics to ensure the best performance. In this paper, we introduce the Copilot evaluation harness: a set of data and tools for evaluating LLM-guided IDE interactions, covering various programming scenarios and languages. We propose our metrics as a more robust and information-dense evaluation than previous state of the art evaluation systems.

We design and compute both static and execution based success metrics for scenarios encompassing a wide range of developer tasks, including code generation from natural language (generate), documentation generation from code (doc), test case generation (test), bug-fixing (fix), and workspace understanding and query resolution (workspace). These success metrics are designed to evaluate the performance of LLMs within a given IDE and its respective parameter space. Our learnings from evaluating three common LLMs using these metrics can inform the development and validation of future scenarios in LLM guided IDEs. 

\end{abstract}






\maketitle

\definecolor{ForestGreen}{RGB}{34,139,34}
\definecolor{RoyalBlue}{RGB}{85,118,209}
\definecolor{DarkPink}{RGB}{255,110,236}

\definecolor{Gray}{gray}{0.9}
\newcommand{\txtmint}[1]{\mintinline[fontsize=\scriptsize, bgcolor=Gray]{text}{#1}}

\newcommand{\ie}{\textit{i.e.,}~}
\newcommand{\eg}{\textit{e.g.,}~}
\newcommand{\etc}{\textit{etc.}~}
\newcommand{\etal}{\textit{et al.}~}

\newcommand{\nb}[2]{
    \fbox{\bfseries\sffamily\scriptsize#1}
    {\sf\small$\blacktriangleright$\textit{#2}$\blacktriangleleft$}
}

\newcommand\MICHELE[1]{\textcolor{blue}{\nb{MICHELE}{#1}}}
\newcommand\ANISHA[1]{\textcolor{ForestGreen}{\nb{ANISHA}{#1}}}
\newcommand\AARON[1]{\textcolor{purple}{\nb{AARON}{#1}}}
\newcommand\SHUBHAM[1]{\textcolor{red}{\nb{SHUBHAM}{#1}}}
\newcommand\YEVHEN[1]{\textcolor{orange}{\nb{YEVHEN}{#1}}}
\newcommand\ROSHANAK[1]{\textcolor{DarkPink}{\nb{ROSHANAK}{#1}}}

\section{Introduction}

The continuous evolution of software development practices has led to a growing interest in the integration of cutting-edge technology to enhance developer productivity \cite{chen2021codex}. One such technology that has garnered considerable attention is the utilization of Large Language Models (LLMs) within Integrated Development Environments (IDEs) \cite{nam2023inide, vscuda}. 
LLMs, exemplified by models like OpenAI's GPT-3.5 \cite{openaigpt35} and GPT-4 \cite{openai2023gpt4}, as well as strong open-source models such as Code Llama \cite{roziere2023code}, offer the promise of acting as intelligent programming assistants. In this paper, we introduce the Copilot evaluation harness for comprehensive exploration of the capabilities and potential applications of LLM-guided programming within the context of IDEs, with a particular focus on their adaptability across diverse programming scenarios and languages.


\begin{figure}[h]
  \includegraphics[width=0.45\textwidth]{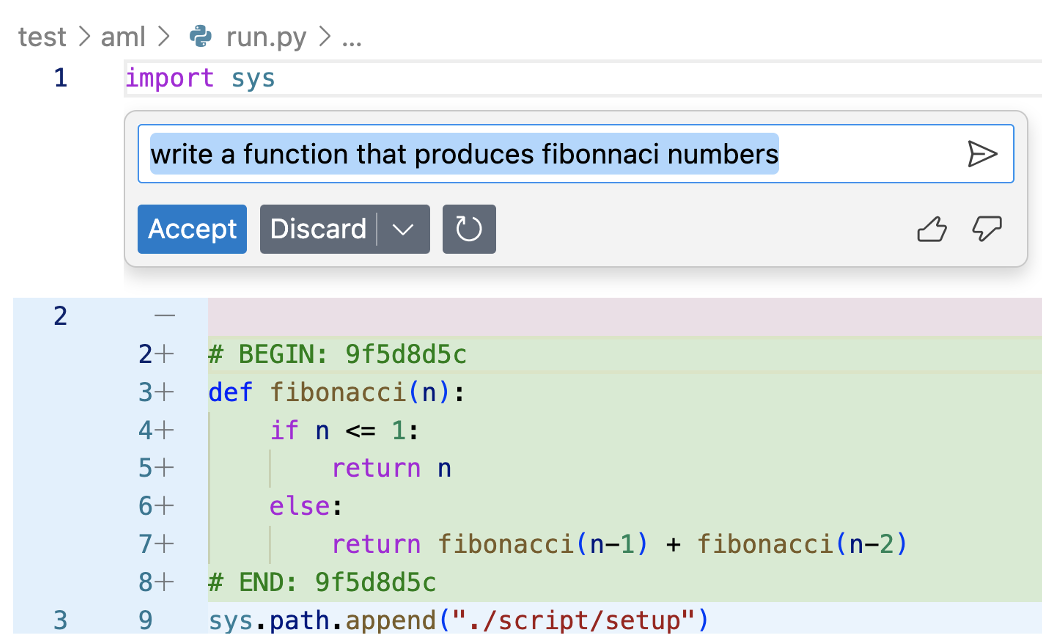}
\caption{A developer has typed the description of a function, which in this case should generate fibonnaci numbers. The LLM has generated the code for this function highlighted in diff format.}
\label{fig:vs-code-generate}
\end{figure}

\begin{figure}[h]
  \includegraphics[width=0.45\textwidth]{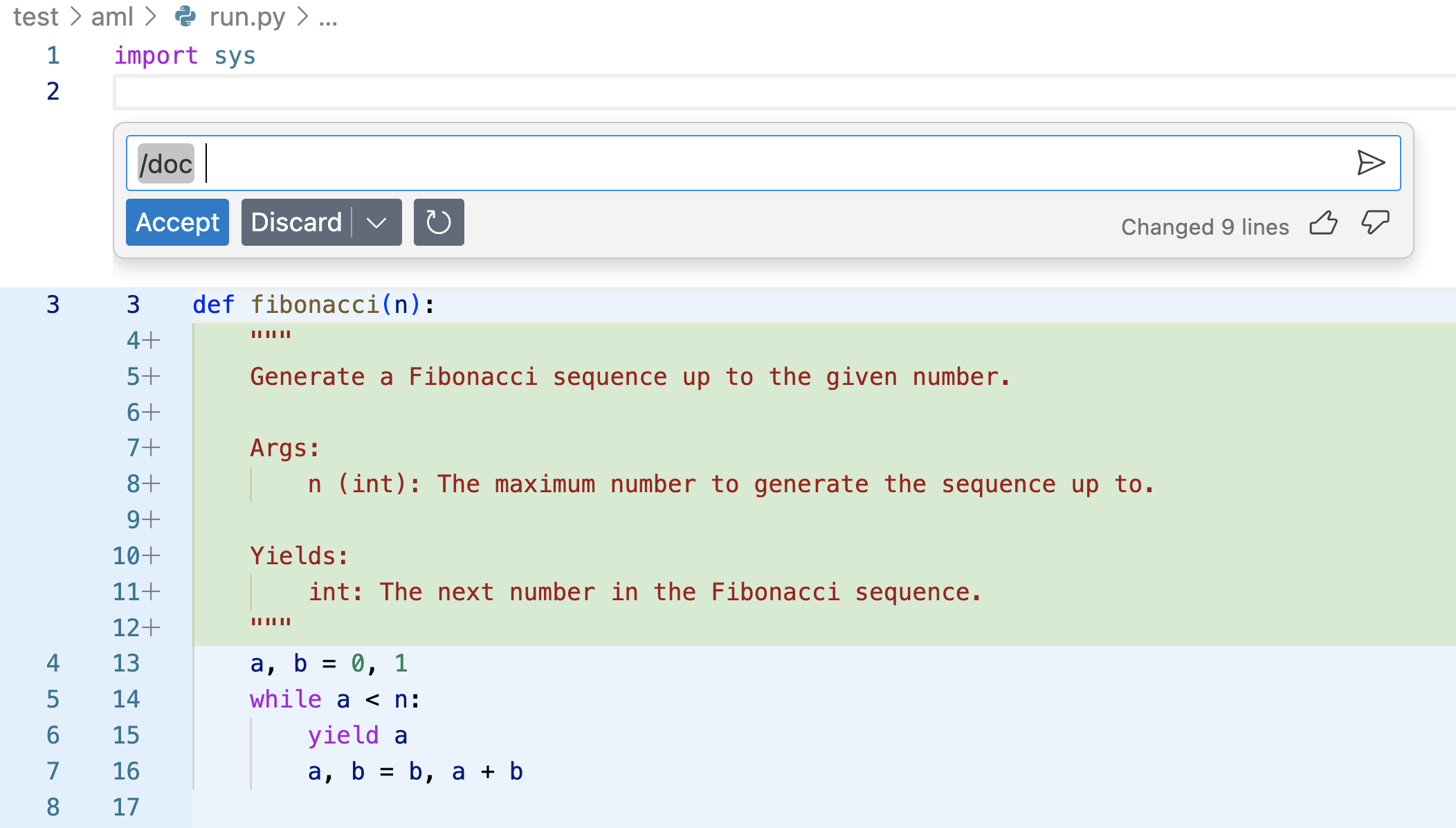}
\caption{A developer uses /doc to generate documentation for a function that generates Fibonacci numbers. The LLM generates the documentation for this function highlighted in diff format.}
\label{fig:vs-code-doc}
\end{figure}

\begin{figure}[h]
  \includegraphics[width=0.45\textwidth]{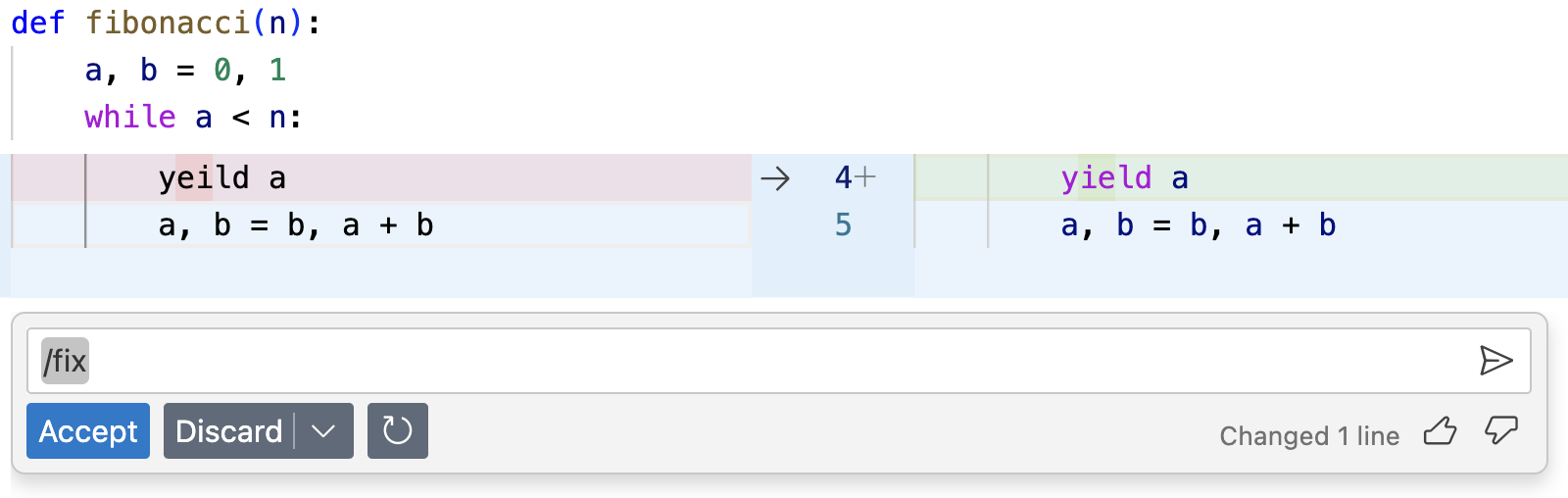}
\caption{A developer asks the model to fix an error in their fibonacci code, and the model presents the fix (spelling the word "yield" correctly) in diff format.}
\label{fig:vs-code-fix}
\end{figure}

At the core of our investigation lies the evaluation of five major software development scenarios. These scenarios encompass a spectrum of developer tasks, each addressing specific challenges and opportunities:

\begin{itemize}
    \item \textbf{Documentation Generation from Code} (doc): LLMs assist in automating the task of generating documentation from code.
    \item \textbf{Bug-Fixing} (fix): LLMs play a crucial role in identifying and rectifying warnings and errors raised by static analysis tools.
    \item \textbf{Code Generation from Natural Language} (generate): LLMs generate code snippets from natural language descriptions.
    \item \textbf{Test Case Generation for Code} (test): LLMs are employed to automatically generate test cases for code, aiming to enhance software quality assurance practices.
    \item \textbf{Workspace Understanding and Query Resolution} (workspace): LLMs help developers understand the current project by responding to developer queries according to a comprehension of the codebase in the local workspace.
\end{itemize}

The metrics in our evaluation harness are designed to evaluate the effectiveness, accuracy, and efficiency of LLM-guided programming interactions across real-world development scenarios. Our framework allows for any IDE to be plugged in and evaluated using our metrics. As such, we provide a system for tuning the IDE parameter space to attain superior LLM-integration outcomes. 

While prior work has offered an evaluation harness for code generation \cite{chen2021codex}, a more comprehensive evaluation harness is needed with the new Large Language Models supporting multiple software engineering scenarios in an IDE. There is a wide parameter space to tune and optimize when integrating an LLM with an IDE: how are the prompts to the model phrased \cite{wei2023chainofthought, sridhar2023hierarchical}? In what order should information be given \cite{liu2023lost}? How are model responses parsed and inserted back into original code? What context should be provided to the model, in addition to the query \cite{petroni2020context, ram2023incontext}? All these factors and more play a role in how well a model can perform within the IDE. 

Previous evaluation harnesses leave gaps in this wide space that we seek to cover with the Copilot Evaluation harness. In the HumanEval dataset \cite{chen2021codex}, for example, models are evaluated on their ability to generate functions from docstrings. This is similar to our method generation evalation metric. However, in HumanEval, the test cases are simple, straightforward, algorithmic coding interview style questions. In our test cases, the model must generate functions from real world code, many of which interact with dozens of other methods and files in order to complete a given task. This level of complexity is necessary to properly evaluate the code generation capabilities of state-of-the-art LLMs in a real world setting. 

Other works use LLMs themselves to evaluate output \cite{zhou2023lima, zhang2023wider, chen2023exploring, fu2023gptscore}. Although this may be effective, there is no guarantee of the model's logic or reasoning, and the natural stochasticity of LLMs makes it difficult to calculate a "true" score for any given test case (i.e. the model may change its answer from run to run). Errors and logic gaps may propagate from the test data into the evaluation results.

With our evaluation framework, we present a new standard of evaluation for model-generated code. Our evaluation harness allows for automatic understanding of how prompt and parameter changes impact performance, across hundreds of test cases spanning a wide range of programming scenarios with real-world code. In this iteration of our work, we discuss detailed results from two of the five metrics outlined above: documentation generation and bug fixing. 

We apply our evaluation framework to evaluate the effectiveness of Visual Studio Code, an IDE used by 15 million programmers across the world. Our evaluation spans a spectrum of LLM models, ranging from proprietary models like OpenAI's GPT-3.5 and GPT-4 to openly available alternatives such as Code Llama. We believe that a diverse set of models is essential to provide a holistic perspective on the capabilities and limitations of LLM-guided programming, catering to the needs and preferences of a wide developer audience.

\section{Related Work}
Below, we explain how our work builds upon and extend the related work on LLMs, Evaluating LLMs, and Evaluating LLMs for software engineering tasks.

\subsection{LLMs}
Large Language Models (LLMs) \cite{devlin2019bert,gao2021making, kombrink2011recurrent} are advanced language models with massive parameter sizes that can understand and generate human language. Many of the well-known LLMs such as GPT-3 \cite{floridi2020gpt}, InstructGPT \cite{ouyang2022training}, and GPT-4 \cite{openai2023gpt4} leverage the Transformer architecture \cite{vaswani2017attention}. In comparison with the traditional machine learning models, LLMs require a large amount of data and very high hardware requirements for training. In return LLMs offer much higher performance than traditional machine learning models when compared their response quality on similar tasks. However, the results from LLMs are less interpretable than the traditional models. 

Building on the success of the LLMs researchers have started to explore the advantages of scaling up LLMs. For example, Gropher \cite{rae2022scaling} has 280 billion parameters, Megatron-turing NLG \cite{smith2022using} has 530 billion parameters and PaLM \cite{chowdhery2022palm} has 540 billion parameters outperforming average humans on the BIGbench benchmark \cite{srivastava2023imitation}. Similarly, researchers also explored fine-tuning LLMs for specific tasks and/or with human feedback \cite{ouyang2022training}. 

In our study, we examine the performance of three prominent LLMs: OpenAI's GPT-3.5, GPT-4, and CodeLlama on five different software engineering scenarios. We have chosen OpenAI's LLMs as representatives of general-purpose language models applicable to Software Engineering tasks, given their large scale and proprietary nature. In contrast, we have included CodeLlama as an illustration of an open-source, smaller, and optimized model fine-tuned specifically for code-related applications.

\subsection{Evaluating LLMs}
Previous work has evaluated the effectiveness of LLMs from various angles including performance in natural language tasks, reasoning, robustness, safety, etc \cite{chang2023survey}. For example, when it comes to sentiment analysis, \cite{liang2023holistic} and \cite{qin2023chatgpt} showed that LLMs perform much better than traditional sentiment analysis models. Similarly, \cite{laskar2023systematic} evaluated ChatGPT's performance on a range of tasks including answering questions, text summarization, code generation, reasoning, and addressing ethical issues. 

Unlike traditional machine learning models where k-fold cross validation was a common evaluation process, LLMs are often evaluated using static data sets. Common dataset for evaluating LLMs include: GLUE \cite{wang2019glue}, SuperGLUE \cite{superglue}, BIGBench \cite{srivastava2023imitation}, Massive Multitask Language Understanding (MMLU) \cite{hendrycks2021measuring}, Ethics Benchmark \cite{hendrycks2023aligning},
and others.

In this paper, we depart from conventional language-based metrics, such as BLEU, commonly employed in previous studies. Instead, we devise metrics tailored specifically for Software Engineering and the tasks under consideration.

\subsection{Evaluating LLMs for Software Engineering Tasks}
LLMs have been widely used in various software engineering tasks, such as code generation, code summarization, code completion, code search, code documentation, code review, bug detection, and software testing. However, evaluating the effectiveness and efficiency of LLMs for SE tasks is not a trivial problem, as there are many factors and challenges involved. In this subsection, we review some of the existing works that have proposed or applied different evaluation methods and metrics for LLMs for SE tasks.

 One of the most comprehensive works is the paper \cite{hou2023large}, which provides a systematic literature review on the intersection of LLMs and SE, covering various aspects such as data collection, preprocessing, application, optimization, evaluation, and prompt engineering. The paper also categorizes and compares different LLMs that have been used in SE tasks, such as GPT-3, CodeBERT, and GraphCodeBERT, and analyzes their strengths and weaknesses. The paper also identifies the current challenges and future directions for LLMs for SE.

CodeXGLUE \cite{lu2021codexglue}, is a comprehensive evaluation platform for LLMs in Software Engineering tasks. CodeXGLUE includes a benchmark dataset with 14 tasks covering code intelligence scenarios and provides baseline models like CodeBERT and CodeGPT. It aims to stimulate research and development in LLMs for SE, offering a diverse dataset for various programming languages and tasks. CodeXGLUE's evaluation metrics, both automatic and human-based, along with a leaderboard and online platform, facilitate fair comparisons between models.

One of the first works that evaluated LLMs for code considering code execution and test cases is the paper \cite{chen2021codex}, which introduces HumanEval, a benchmark dataset and a challenge for measuring the functional correctness of LLMs trained on code. HumanEval consists of 164 hand-written programming problems in Python, each with a function signature, a docstring, a body, and several unit tests. The problems cover various topics, such as language comprehension, algorithms, and simple mathematics, and some of them are comparable to simple software interview questions. The goal of HumanEval is to measure the ability of LLMs to synthesize programs from docstrings that pass the given test cases.

In our research, we build upon the foundation laid by prior works in the literature, seeking to enhance their contributions. Like HumanEval, we incorporate considerations of code execution and test cases, but we extend both in terms of the breadth of SE tasks addressed and the refinement of evaluation metrics. In contrast to HumanEval, our evaluation encompasses large and real-world codebases. Furthermore, our emphasis is on developing a comprehensive evaluation framework for LLM-guided programming within IDE interactions, with a particular focus on their practicality across diverse programming languages and scenarios.

\section{Evaluating LLM-Guided Software Programming}
In addition to HumanEval \cite{chen2021codex}, match-based metrics such as BLEU \cite{papineni2002bleu} or Code-BLEU \cite{ren2020codebleu} are commonly adopted to benchmark LLM's performance in software engineering tasks. As LLMs become more ubiquitous and powerful, many researches use LLM models themselves to evaluate LLM output \cite{zhou2023lima, zhang2023wider, chen2023exploring, fu2023gptscore}. However, previous work suggests alternative metrics such as functional correctness better reflect success of generative models in code generation \cite{chen2021codex}, code translation\cite{roziere2020}, and other tasks. Building upon previous work in this area, we expand the HumanEval harness and evaluate IDE-integrated model competency in the five software engineering tasks listed above.   

\subsection{Documentation Generation from Code (doc)} 
This task involves generating documentation for a method. Figure \ref{fig:vs-code-doc} shows an example in VS Code IDE. In this case the developer asks the LLM to generate a documentation for a Fibonacci function using /doc.  

\subsubsection{Metrics} 
In this scenario, we consider a docstring generation to be successful if the location, format, and coverage of the generated text is correct. We report the following metrics for this scenario: 

\begin{itemize} 
    \item \emph{Syntax Correctness}: We check that the docstring has been inserted into the code in such a way that it does not disrupt the syntax of the file with its addition.  
    \item \emph{Format Correctness}: If the documentation comment is placed in a syntactically acceptable manner for the given language, we further check for the correctness of documenting the return statement, function arguments with their types, function name, and whether a function description was written.  
\end{itemize} 

\subsubsection{Evaluation Procedure} 
We begin with a set of methods. For each method, we provide the method's signature and body to the LLM as context.  We then prompt the LLM with a request to generate documentation for the method, and return the input function with the generated docstring inserted in the correct location within the function .  

After the LLM generates the documentation and the generated docstring is inserted into the code file, we evaluate the syntax correctness of the file with the generated docstring, as well as the correctness of the docstring itself.

\subsection{Bug-Fixing (fix)} 

This task involves using LLMs to fix bugs identified by static analysis tools, with an expectation that the resulting fixed code will have fewer errors overall than the original code. We use the following static analyzers: 

\begin{itemize} 
\item javascript: eslint \cite{eslint};  
\item ts: eslint \cite{eslint}, tsc (typescript compiler); 
\item python: pylint \cite{pylint}, pyright \cite{pyright}; 
\item java: spotbugs \cite{spotbugs}; 
\item c\#: roslyn \cite{roslyn}; 
\item cpp: clang \cite{clang}. 
\end{itemize} 

If the original error is fixed but another error is introduced in its place, the test case will fail. 

Figure \ref{fig:vs-code-fix} shows an example in the VS Code IDE. A programmer has an error because of a misspelling of the word "yield", and the model corrects the error. 

\subsubsection{Metrics} 
In this scenario, we consider a bug fix to be successful if the resulting code is syntactically correct and the corresponding static analysis warning or error has disappeared. 

\begin{itemize} 
    \item \emph{Syntax Correctness}: we confirm that the code file with the bug fix remains syntactically correct. 
    \item \emph{Fix Rate}: we check that an existing static analysis warning or error in the code has been successfully resolved by the suggested changes, without introducing any other errors. 
\end{itemize} 

\subsubsection{Evaluation Procedure} 
Given a set of bugs found by static analyzer tools, we provide the file contents and diagnostic information to the LLM to generate a fix. We assess whether the model fixed the original error, whether it created any new errors, and whether the model-modified code remained syntactically correct after the fix was inserted. 

\subsection{Code Generation from Natural Language (generate)}
This task involves generating a code snippet from a natural language description. Figure \ref{fig:vs-code-generate} shows an example of such a task in the VS Code IDE. In this case, the developer asks the LLM to write a function that produces the first $n$ values of the Fibonacci sequence, and the editor shows the generated function in a diff view. 
\begin{figure*}
  \includegraphics[width=\textwidth]{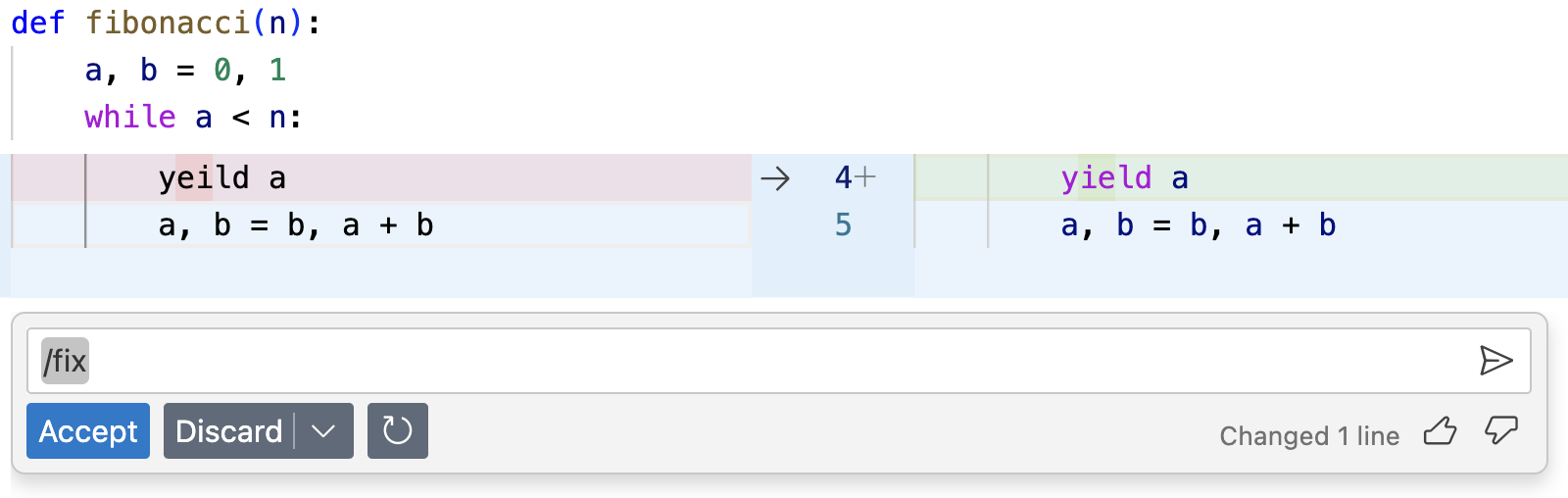}
\caption{A developer asks the model to fix an error in their fibonacci code, and the model presents the fix (spelling the word "yield" correctly) in diff format.}
\label{fig:vs-code-fix}
\end{figure*}
\subsubsection{Metrics}
Similar to previous evaluations of code generations \cite{chen2021codex}, we consider a generated code snippet to be successful if the generated code is syntactically correct and all test cases covering the generated code pass. Therefore, we report the following metrics for this scenario: 

\begin{itemize}
    \item \emph{Syntax Correctness}: We compute and report the percentage of generated code that is syntactically correct. For this metric, we check the syntax correctness of the generated code using a language-specific parser (e.g., tree-sitter for each language). 
    \item \emph{Test Pass Rate}: We check the number of passing and failing tests and compute the passing test ratio. To compute this number, we execute the entire test suite of the user project and track which tests fail that passed prior to the model's code injection. 
\end{itemize}

\subsubsection{Evaluation Procedure}
We begin with a set of repositories with test cases. From each repository, we select the methods that are: 1) covered by the test cases in the given repository's test suite, and 2) have a docstring. For each method, we ask an LLM to generate the body of the method given the method's signature and docstring. We provide the contents of method's file as context to the LLM, replacing the original method body with a commented line reading "Your Code Here." 
 
After the LLM generates the method body, we put the generated code back in place of the original method body and evaluate the code by running the repository's test suite against the new method body. We then compute and report the syntax correctness and test pass rate, as explained above.

\subsection{Test Case Generation for Code (test)}
\begin{figure*}
  \includegraphics[width=\textwidth]{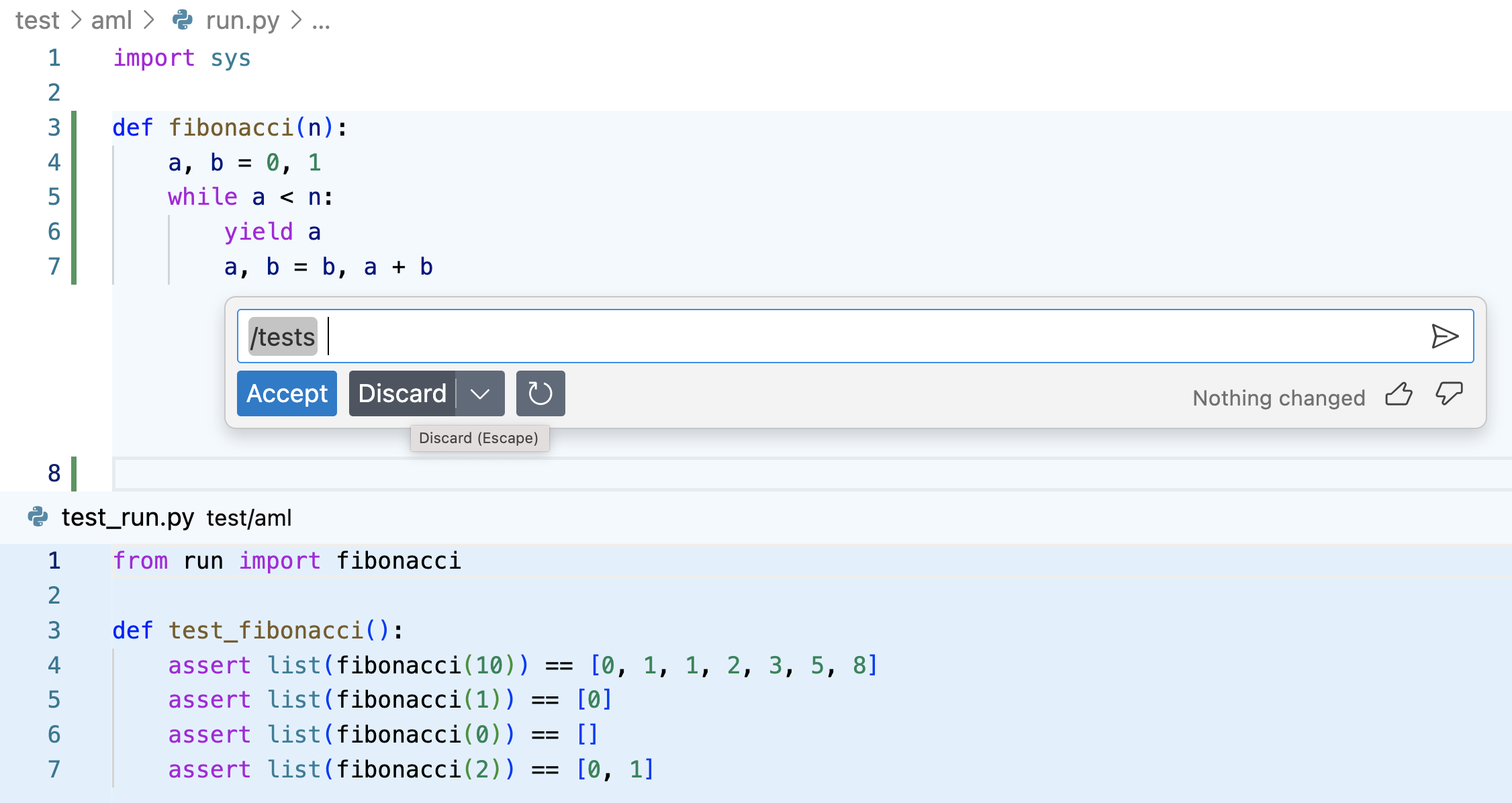}
\caption{A developer uses /test to generate a test for a function that generates Fibonacci numbers. The LLM generates the test\_fibonacci function for this function in a test file.}
\label{fig:vs-code-test}
\end{figure*}

This task involves using LLMs to generate test cases for code. Developers usually shortcut when it comes to writing unit tests. Automating test generation can motivate more developers to include unit tests. Figure \ref{fig:vs-code-test} shows an example of a developer requesting tests in the VS Code IDE. In the example case, the developer asks the LLM to generate a test for a Fibonacci function using the /test chat scenario command. 

\subsubsection{Metrics}
In this scenario, we consider a generated test to be successful if it is syntactically correct and can pass on execution. Note that, for this evaluation, this means we assume the code for which the test was written is correct.
\begin{itemize}
    \item \emph{Syntax Correctness}: We compute the percentage of generated tests that are syntactically correct. We check the syntax correctness of the generated tests using a language-specific parser. 
    \item \emph{Generated Test Pass Rate}: We compute the pass rate of the generated test. We assume the original method is correct, and execute the generated test on its focal method. 
\end{itemize}
\subsubsection{Evaluation Procedure}
Given a set of methods, we provide the method signature, docstring, and body as context to the LLM to generate a test for each focal method. 

Once the LLM generates a test for the method, we add the test to the repository containing the method, and attempt to execute the test.


For Javascript and Typescript, we generate tests using either the Jest or Mocha library. The original test suite of the repository does not need to be written with either library, but each method's original file must be able to pass without errors when a trivial test case (which essentially just asserts true) is appended to the file. When evaluating the generated tests, we temporarily append them to the focal method's file to mitigate import errors, and run the entire file. If running the file with a trivial test case appended (e.g. a test that should always be true) returns false or an error, we know the results from the generated test on that file are not reliable.

\subsection{Workspace Understanding and Query Resolution (workspace)}
In the Workspace task, we give the model a user's natural language query, and ask it to identify relevant snippets of the codebase that may aid in answering the user's question. This tests a model's ability to comprehend both natural language requests from a user and large amounts of code. 

\subsubsection{Metrics}
We evaluate the quality of an LLM's retrieved snippets in two ways:
\begin{itemize}
    \item Mean Reciprocal Rank (MRR): Given a ranked list of the model's retrieved snippets, we calculate $\frac{1}{r}$, where $r $ is the rank of the correct snippet in the model's list. So, if the model ranks the correct snippet second, we would consider the model's score for that test case to be $\frac{1}{2}$. MRR is the mean of all the test case scores.
    \item End to End Keyword Detection: We begin with a manually created dataset of user queries and keywords associated with the correct answer to the query. We take the model's ranked list of retrieved snippets and pass it to the model along with each user query. Then, we detect whether or not the associated keyword appeared in the model's response, given both the query and the retrieved results. 
\end{itemize}

\subsubsection{Evaluation Procedure}
For each datapoint, we provide the LLM with a user query and the full context of the codebase associated with the given query. We ask the LLM to retrieve a ranked list of relevant code snippets from the codebase. We directly evaluate the quality of the model's retrieved results using MRR, a metric that scores how well models are able to find the most relevant code snippets during retrieval.\\
We also evaluate the quality of all the retrieved code snippets by asking the model to answer the original user query, providing the query and the snippets as context. We search the model's final response for a set of keywords associated with the given query to determine whether or not the model was able to find the information it needed to fully answer the question.\\
With this metric, we evaluate the model's retrieval abilities on an end to end scale, and determine a model's skill at finding code snippets that would actually help it answer the question at hand.

\section{Copilot Evaluation Harness}
We introduce the end to end Copilot Evaluation Harness for computing evaluation metrics as described above. First, we share the details of collecting the data required for each evaluation. Then, we explain the process of creating a test environment given each language and the need to build and run tests. Finally, we give additional specific implementation details about the evaluation process for each metric.

\subsection{Data Collection} 

Our dataset is made up of methods from hundreds of public GitHub repositories across 6 languages: JavaScript, Typescript, Python, Java, C/C++, and C\#.  Some of our evaluations require the ability to build and run tests for repositories associated with test cases. To meet this requirement, we have developed a build agent as part of our evaluation harness that attempts various build and test strategies on any arbitrary repository. In addition, we have the capability to run static analysis tools on the repositories that we can build and test. This build agent is essential in collecting the test datasets and performing evaluations. 

For each language, we sample from Github public repositories whose code we are able to build and whose test suites we are able to run using our build agent. The build agent supports Node 18+, Python 3.8+, Java JDK 1.8 (requiring Maven), .NET 6.0, 7.0 and 8.0, and a manually curated set of C++ repositories. We resorted to manually gathering C++ repositories due to the wide variability of C++ build steps. We ignore repositories that are smaller than 1 MB and larger than 100 MB. We ignore repositories that take longer than 10 minutes to build and run tests. Lastly, we ignore repositories that do not contain any methods. 

\subsubsection{Javascript and Typescript} 
In Javascript and Typescript, we sub-select on repos that contain a $package.json$ file at the root directory. The $package.json$ file works in concordance with npm (Node Package Manager) to handle various tasks within the repo, such as specifying dependencies for installation and running the test suite. We rely on npm for our evaluation of Javascript and Typescript code, so we only consider repos whose infrastructure is built to be managed with npm.

\subsubsection{Java}
In Java, we consider repositories that leverage Maven for their build process. In addition, as of writing, we only consider projects that use JDK 1.8.

\subsubsection{Python}
In Python, we only consider repositories for which we are able to successfully install all dependencies within a virtual environment.

\subsubsection{C/C++}
In C/C++, we leverage clang for building projects. Because of the sheer variety of ways that C/C++ repositories can  be built, we present a set of manually curated repositories which we have verified will build and test within a docker image.

\subsection{Test Case Collection}

After identifying suitable repositories for each language, we generate test cases for each evaluation metric based on the code within the repositories. Most evaluations require identifying methods that meet certain conditions, such as being covered by existing tests or containing a warning from a static analysis tool. The criteria for generating evaluation test cases varies from metric to metric, and is explained for each metric below. 

\subsubsection{Documentation Generation from Code (doc)}
We create test cases by identifying methods in the repository that are longer than three lines and are not a result of minification or obfuscation. We provide the method and ask the coding assistant being evaluated to generate a docstring for the method. We consider a docstring generation to be successful if the location, format, and coverage of the generated text is correct. 

\subsubsection{Bug Fixing (fix)}
We create test cases based on static analysis tool warnings and errors flagged on a given repository. We only consider static analysis warnings that are not related to imports or configuration because such issues are difficult to fix with only a single file as context. We consider a generated fix to be successful if it is syntactically correct and strictly reduces the numbers of static analysis warnings on execution. We must consider a strict decrease rather than the presence of the original warning or error, because it is possible for the coding assistant to fix the original issue while introducing a new issue, which a developer would not look upon as a complete fix.

\subsubsection{Code Generation from Natural Language (generate)}
We create test cases by identifying methods in a given repository that are covered by some existing passing test. The test case gives the coding assistant visibility of the entire file up to and including the method signature. The coding assistant is then asked to generate the method body associated with the method signature. We consider a generated code snippet to be successful if the generated code is syntactically correct and all test cases covering the generated code pass.

\subsubsection{Test Generation from Code (test)}
We create test cases by identifying methods within a given repository. We ask the coding assistant to provide a working test for the given method. We consider the generated test to be successful if it invokes the given method and passes execution.

\subsubsection{Workspace Understanding and Query Resolution (workspace)}
We collected questions from developers about certain aspects of their project workspace, such as the idiomatic way to build a certain feature. The context fetching that takes place as a part of the workspace command will return several related code snippets. We evaluate the quality of an LLM’s retrieved snippets using MRR as explained above.

\section{Experiments}
Using Copilot evaluation harness metrics and the test cases detailed above, we compute the success of two OpenAI models: GPT-3.5 and GPT-4, as well as CodeLlama on the document generation and bug fixing scenarios using the an LLM powered chat extension in VSCode IDE with more than 700K active users as the code assistant. 

Our experimentation aims to answer the following research questions:
\begin{itemize}
\item \textbf{RQ1. Model Comparison}: How do different LLM's compare to one another when integrated with a coding assistant?
\item \textbf{RQ2. Integration Improvements}: What insights can the Copilot Evaluation harness provide engineers to improve the integration of LLM in a coding assistant?
\item \textbf{RQ3. Data Validity}: How do our evaluation test cases compare with actual usage of a LLM powered coding assistant? Do the test cases in our harness reflect how real-world users interact with a LLM powered coding assistant?
\end{itemize}

In this section, we discuss findings pertaining to these research questions. 
\begin{table}[h]
    \centering
    \begin{tabular}{@{}llcc@{}}
 \multicolumn{4}{c}{\textbf{Doc}}\\
\toprule
\textbf{Language}&\textbf{Model}&\textbf{Syntax}&\textbf{Format}\\
&&\textbf{Correctness}&\textbf{Correctness}\\
\toprule
\multirow{3}{*}{Python}&  GPT-4 & 100\% & 83\%\\&  GPT-3.5 & 100\% & 87\%\\&CodeLlama & 100\% & 87\%\\
\toprule
\multirow{3}{*}{Javascript}&  GPT-4 & 83\% & 100\%\\&  GPT-3.5 & 83\% & 100\%\\&CodeLlama & 79\% & 55\%\\
\toprule
\multirow{3}{*}{Typescript}&  GPT-4& 96\% & 79\%\\&  GPT-3.5 & 96\% & 86\%\\&CodeLlama & 77\% & 65\%\\
\toprule
\multirow{3}{*}{Java}&  GPT-4 & 100\% & 93\%\\&  GPT-3.5 & 100\% & 80\%\\&CodeLlama & 100\% & 64\%\\
\toprule
\multirow{3}{*}{C\#}&  GPT-4 & 100\% &89\% \\&  GPT-3.5  & 100\% & 75\%\\&CodeLlama & 94\% & 67\%\\
\toprule
\multirow{3}{*}{C/C++}&  GPT-4 & 92\% & 94\%\\&  GPT-3.5 & 92\% & 77\%\\&CodeLlama & 90\% & 38\%\\
\bottomrule
\end{tabular}
\caption{LLMs performance on the Doc chat scenario across Python, Javascript, Typescript, Java, C\# and C/C++ for the specific success metrics of the Doc scenario.}
\label{tab:doc_results}
\end{table}

\begin{table}[h]
    \centering
\begin{tabular}{@{}llcc@{}}
 \multicolumn{4}{c}{\textbf{Fix}}\\
\toprule
\textbf{Language}&\textbf{Model}&\textbf{Syntax }&\textbf{Bugs}\\
&&\textbf{Correctness}&\textbf{Fixed}\\
\toprule
\multirow{3}{*}{Python}&  GPT-4 & 96\% & 74\%\\&  GPT-3.5& 93\% & 68\%\\&CodeLlama & 88\% & 39\%\\
\toprule
\multirow{3}{*}{Javascript}&  GPT-4  & 92\% & 81\%\\&  GPT-3.5 & 85\% & 74\%\\&CodeLlama & 39\% & 26\%\\
\toprule
\multirow{3}{*}{Typescript}&  GPT-4 & 83\% &75\% \\&  GPT-3.5 & 74\% & 75\%\\&CodeLlama & 70\% & 30\%\\
\toprule
\multirow{3}{*}{C\#}&  GPT-4 & 98\% & 58\%\\&  GPT-3.5  & 96\% & 65\%\\&CodeLlama  & 84\% & 50\%\\
\bottomrule

\end{tabular}

\caption{LLMs performance on the Fix chat scenario across Python, Javascript, Typescript and C\# for the specific Fix success metrics. }
\label{tab:fix_results}
\end{table}
\subsection{RQ1. Model Comparison}
Below we discuss our learnings comparing three state of the art LLMs when used to power our target chat extension in VSCode. 
\subsubsection{Documentation Generation from Code (doc)}
Table \ref{tab:doc_results} shows that, for docstring generation, GPT-4 generally outperforms GPT-3.5 and Code Llama. GPT-3.5 and GPT-4 are very similar in performance to one another, with Code Llama slightly behind. The key exceptions here are Python, where Code Llama performs at a slightly higher level than GPT-4, and C/C++, where Code Llama performs significantly worse. One possible explanation is that 
GPT-3.5 and GPT-4 were trained on a massive corpus, encompassing much of the open source code on the internet. As such, the GPT models' performance could be inflated by the fact that it has seen many different code patterns. 
Code Llama, a comparatively small model, is much less likely to have seen a given code snippet, potentially hindering its performance in comparison to the GPT models.


\begin{figure}[H]
\vspace{-0.0cm}
    \centering
\begin{adjustbox}{width=0.5\textwidth}
\begin{tabular}{c}
\toprule
  Passing Fix Request\\
\midrule
\begin{minipage}[t]{0.45\textwidth}
\begin{minted}{text}
You have been given the file contents of  lightning_train_net.py. 
The following code snippet within the file has a bug:
\end{minted}
\begin{minted}[escapeinside=||]{python}
def training_epoch_end(self, training_step_outputs):
    self.iteration_timer.after_train()
    if comm.is_main_process():
        self.checkpointer.save("model_final")
    for writer in self.writers:
        writer.write()
        writer.close()
    self.storage.__exit__(None, None, None)
\end{minted}
\begin{minted}{text}
This is the line with the error:
\end{minted}
\begin{minted}[escapeinside=||]{python}
for writer in self.writers:
\end{minted}
\begin{minted}{text}
This is the problem with the line:
\end{minted}
\begin{minted}[escapeinside=||]{python}
Object of type "None" cannot be used as iterable.
\end{minted}
\begin{minted}{text}
Describe in a single sentence how you would solve the 
problem. Then, fix the error.
\end{minted}
\end{minipage}\\
\midrule
GPT-4 Passed Response \\
\midrule
\begin{minipage}[t]{0.45\textwidth}
\begin{minted}{text}
To solve the problem, I would add a condition to check 
if `self.writers` is not None before iterating over it.
\end{minted}
\begin{minted}[escapeinside=||]{python}
def training_epoch_end(self, training_step_outputs):
self.iteration_timer.after_train()
if comm.is_main_process():
    self.checkpointer.save("model_final")
if self.writers is not None:
    for writer in self.writers:
        writer.write()
        writer.close()
self.storage.__exit__(None, None, None)

\end{minted}
\end{minipage} \\
\\
\bottomrule
\end{tabular}
\end{adjustbox}
\vspace{-0.0cm}
\caption{Example prompt and response for a passing fix test case.}
\vspace{-0.0cm}
\label{fig:fix_pass}
\end{figure}

\begin{figure}[H]
\vspace{-0.0cm}
    \centering
\begin{adjustbox}{width=0.5\textwidth}
\begin{tabular}{c}
\toprule
  Failed Fix Request\\
\midrule
\begin{minipage}[t]{0.45\textwidth}
\begin{minted}{text}
You have been given the file contents of tutorial009.py.
The following code snippet within the file has a bug:
\end{minted}
\begin{minted}[escapeinside=||]{python}
def select_heroes():
    with Session(engine) as session:
        statement = select(Hero).where(or_(
        Hero.age <= 35, Hero.age > 90
        ))
        results = session.exec(statement)
        for hero in results:
            print(hero)

def main():
    create_db_and_tables()
    create_heroes()
    select_heroes()
\end{minted}
\begin{minted}{text}
This is the line with the error:
\end{minted}
\begin{minted}[escapeinside=||]{python}
statement = select(Hero).where(or_(
            Hero.age <= 35, Hero.age > 90
            ))
\end{minted}
\begin{minted}{text}
This is the problem with the line:
\end{minted}
\begin{minted}[escapeinside=||]{python}
Operator ">" not supported for "None".
\end{minted}
\begin{minted}{text}
Describe in a single sentence how you would solve the 
problem. Then, fix the error.
\end{minted}
\end{minipage}\\
\midrule
GPT-4 Failed Response \\
\midrule
\begin{minipage}[t]{0.45\textwidth}
\begin{minted}{text}
The problem arises because the `age` field of the `Hero` 
model is optional and can be `None`, and the 
comparison operator `>` is not supported for `None`. 
To fix this, we need to add a condition to check if 
`Hero.age` is not `None` before comparing it with a 
number.
\end{minted}
\begin{minted}[escapeinside=||]{python}
def select_heroes():
    with Session(engine) as session:
        statement = select(Hero).where(or_(
        Hero.age <= 35, Hero.age > 90, Hero.age.isnot(None)
        ))
        results = session.exec(statement)
        for hero in results:
            print(hero)


def main():
    create_db_and_tables()
    create_heroes()
    select_heroes()
\end{minted}
\end{minipage} \\
\\
\bottomrule
\end{tabular}
\end{adjustbox}
\vspace{-0.0cm}
\caption{Example prompt and response for the failed fix chat scenario. Here, the model determines what the problem might be and attempts to fix it. However, the fix is not correct, and the same syntax error still exists, because it tries to do the > operation before checking whether the age is None.}
\vspace{-0.0cm}
\label{fig:fix_fail}
\end{figure}
\subsubsection{Bug-Fixing (fix)}

Table \ref{tab:fix_results} shows the results for bug fixing: similarly to the docstring generation evaluation, GPT-4 tends to slightly outperform GPT-3.5, with Code Llama further behind. For bug fixing, the exception seems to be C\#, for which all three models seem to struggle, with GPT-3.5 ultimately outperforming both GPT-4 and Code Llama.
Figures \ref{fig:fix_pass} and \ref{fig:fix_fail} show a passed and failed example with GPT-4. In Figure \ref{fig:fix_pass}, the model adds an $if$ statement to check whether $self.writers$ is None before trying to use it as an iterable, which solves the error. In Figure \ref{fig:fix_fail}, the model similarly adds a check for whether $Hero.age$ is None. However, since it adds the check after the operation that is causing the error, the error continues to occur. Although the model was able to identify a potential fix, it inserted the check in the wrong location and was not able to fix the bug. 

\begin{figure}[H]
\vspace{-0.0cm}
    \centering
\begin{adjustbox}{width=0.5\textwidth}
\begin{tabular}{c}
\toprule
  GPT-3.5 vs. GPT-4 Fix Request Response\\
\midrule
\begin{minipage}[t]{0.45\textwidth}
\begin{minted}{text}
You have been given the file contents of picgoCoreIPC.ts
The following code snippet within the file has a bug:
\end{minted}
\begin{minted}[escapeinside=||]{typescript}
  const handler = (msg: string) => {
    if (msg === 'NPM is not installed') {
      dialog.showMessageBox({
        title: T('TIPS_ERROR'),
        message: T('TIPS_INSTALL_NODE_AND_RELOAD_PICGO'),
        buttons: ['Yes']
      }).then((res) => {
        if (res.response === 0) {
          shell.openExternal('https://nodejs.org/')
        }})}}
\end{minted}
\begin{minted}{text}
This is the line with the error:
\end{minted}
\begin{minted}[escapeinside=||]{typescript}
      }).then((res) => {
\end{minted}
\begin{minted}{text}
This is the problem with the line:
Parameter 'res' implicitly has an 'any' type.
Describe in a single sentence how you would solve the problem. 
Then, fix the error.
\end{minted}
\end{minipage}\\
\midrule
GPT-3.5 Passing Response \\
\midrule
\begin{minipage}[t]{0.45\textwidth}
\begin{minted}{text}
To fix the issue of the implicitly typed parameter 'res', you can 
explicitly type it as 'any' in the arrow function
\end{minted}
\begin{minted}[escapeinside=||]{typescript}
  const handler = (msg: string) => {
    if (msg === 'NPM is not installed') {
      dialog.showMessageBox({
        title: T('TIPS_ERROR'),
        message: T('TIPS_INSTALL_NODE_AND_RELOAD_PICGO'),
        buttons: ['Yes']
      }).then((res: any) => { // Explicitly type 'res' as 'any'
        if (res.response === 0) {
          shell.openExternal('https://nodejs.org/')
        }})}}
\end{minted}
\end{minipage} \\
\midrule
GPT-4 Failing Response \\
\midrule
\begin{minipage}[t]{0.45\textwidth}
\begin{minted}{text}
To solve the problem, I would explicitly define the type of 'res' 
in the promise callback
\end{minted}
\begin{minted}[escapeinside=||]{typescript}
  const handler = (msg: string) => {
    if (msg === 'NPM is not installed') {
      dialog.showMessageBox({
        title: T('TIPS_ERROR'),
        message: T('TIPS_INSTALL_NODE_AND_RELOAD_PICGO'),
        buttons: ['Yes']
      }).then((res: Electron.MessageBoxReturnValue) => {
        if (res.response === 0) {
          shell.openExternal('https://nodejs.org/')
        }})}}
\end{minted}
\end{minipage} \\
\bottomrule
\end{tabular}
\end{adjustbox}
\vspace{-0.0cm}
\caption{Example prompt and response for the same test case, where both GPT-3.5 and GPT-4 should fail, but only GPT-4 fails because it attempts a more nuanced approach to fixing the bug than GPT-3.5.}
\vspace{-0.0cm}
\label{fig:case-1924}
\end{figure}

A common cause of differing results between the GPT-3.5 and GPT-4 LLMs occurs when the models try to resolve the "has an 'any' type" error, as can be seen in Figure \ref{fig:case-1924}. When the GPT-4 model attempts to specify the type of a variable named $res$, it predicts the type as $Electron.MessageBoxReturnValue$ and casts the variable to that type. That type, however, is not a valid return type for the code. GPT-3.5, on the other hand, casts the variable to type $any$, thereby circumventing more complex issues, but leaving in a code smell (since it is not good practice to cast variables to type $any$). In this case, GPT-3.5 passes our evaluation, while GPT-4 fails, even though GPT-4's attempted fix is more nuanced and advanced. Upon closer inspection of the cases where GPT-3.5 succeeds and GPT-4 fails, we see this phenomenon frequently: GPT-4 fails with a more complicated approach, while GPT-3.5 technically passes, but with a rudimentary and sub-optimal solution.

\subsection{RQ2. Integration Improvements}
Below we discuss how our evaluation harness can be used to learn insights on how to better integrate LLMs with IDEs.
\subsubsection{Documentation Generation from Code (doc)}
Further inspection of our results in table \ref{tab:doc_results} reveals four classes of errors that cause the docstring generation evaluation to fail:
\begin{enumerate}
\item \textbf{Code Logic Changes:} The model changes the fundamental logic of the code when it rewrites the focal function to the file along with the generated docstring.
\item \textbf{Syntax Changes:} The model changes the syntax of the focal code when writing to the file. This includes changes such as adding semicolons to the end of lines, or adding type decorators to the function signature.
\item \textbf{Incomplete Docstrings:} The model generates a description for the correct function, but does not describe the returned object and every parameter of the function. 
\item \textbf{Irrelevant Docstrings:} The model returns a docstring that does not pertain to the code block we asked it to document.
\end{enumerate}

Upon closer inspection, in cases where only one of GPT-3.5 or GPT-4 passes, we notice that the GPT-4 model is more likely to make changes to the focal code that make the code cleaner than GPT-3.5. For example, in Figure \ref{fig:doc_case315}, the GPT-4 model adds type decorators to the inputs of the function, and specifies the return type. Although the GPT-4 model's docstring is correct (and more detailed than that of the GPT-3.5 model), the GPT-4 model fails this test case, since we expect the model to make no changes to the focal code. However, such an error shows how the GPT-4 model's attempts at a more involved improvement may decrease its score without being indicative of worse performance.

Based on this finding, we inserted an additional instruction in the coding assistant's docstring generation prompt that tells the model specifically not to change any of the focal code. This resulted in a significant improvement in the evaluation results for all the languages, ranging from 5\% in C++ to 11\% in Java. 

\begin{figure}[H]
\vspace{-0.0cm}
    \centering
\begin{adjustbox}{width=0.5\textwidth}
\begin{tabular}{c}
\toprule
  Doc: GPT-4 Syntactical Change \\
\midrule
\begin{minipage}[t]{0.45\textwidth}
\begin{minted}{text}
Write documentation for dump
-----------------------------
\end{minted}
\begin{minted}[escapeinside=||]{typescript}
function dump(classFunction, pref) {
  window.document.write("<b>" + pref + classFunction.name 
                            + "</b> <br/>");
  const keys = Object.keys(classFunction);
  if (keys.length > 0 && keys[0] !== "0") {
    for (const prop of keys) {
      dump(classFunction[prop], pref + 
                            classFunction.name + ".");
    }}}
\end{minted}
\end{minipage}\\
\midrule
GPT-3.5 Passing Response \\
\midrule
\begin{minipage}[t]{0.45\textwidth}

\begin{minted}[escapeinside=||]{typescript}
/**
 * Recursively dumps the properties of a class or object.
 * 
 * @param classFunction - The class or object to dump.
 * @param pref - The prefix to use for indentation.
 */
function dump(classFunction, pref) {
  window.document.write("<b>" + pref + classFunction.name 
                            + "</b> <br/>");
  const keys = Object.keys(classFunction);
  if (keys.length > 0 && keys[0] !== "0") {
    for (const prop of keys) {
      dump(classFunction[prop], pref + 
                            classFunction.name + ".");
    }}}
\end{minted}
\end{minipage} \\
\midrule
GPT-4 Failing Response \\
\midrule
\begin{minipage}[t]{0.45\textwidth}

\begin{minted}[escapeinside=||]{typescript}
/**
 * Writes the name of the given class function and its 
 * properties to the document.
 * If the class function has properties that are also functions, 
 * it recursively writes their names and properties as well.
 *
 * @param classFunction - The function to be dumped. 
 *               It should be a class or a function.
 * @param pref - The prefix to be added before the function name.
 *         It is used for nested functions to show the hierarchy.
 */
function dump(classFunction: Function, pref: string): void {
  window.document.write("<b>" + pref + classFunction.name 
                            + "</b> <br/>");
  const keys = Object.keys(classFunction);
  if (keys.length > 0 && keys[0] !== "0") {
    for (const prop of keys) {
      dump(classFunction[prop], pref + 
                            classFunction.name + ".");
    }}}
\end{minted}
\end{minipage} \\

\\
\bottomrule
\end{tabular}
\end{adjustbox}
\vspace{-0.0cm}
\caption{Example prompt and response for doc evaluation. GPT-3.5 passed, and GPT-4 failed because it returned the original function with parameter and return types. This fails our evaluation because we require the model to leave the focal code unchanged.}
\vspace{-0.0cm}
\label{fig:doc_case315}
\end{figure}

 We also see that GPT-4 is better at following specific instructions. This is highlighted by the example in Figure \ref{fig:doc_case5401}, where GPT-4 generates documentation for the Vec class instead of the $get\textunderscore colour\textunderscore at$ function. At first, this appears to be an error on GPT-4's part. However, a review of the prompt shows an incorrect parsing of the function name for C++ erroneously puts Vec in the prompt rather than $get\textunderscore colour\textunderscore at$. Even though GPT-3.5 received the same instructions, it wrote documentation for $get\textunderscore colour\textunderscore at$. From examples such as this, we see that the GPT-4 model is more sensitive to instructions, and we must design our prompts accordingly.

\begin{figure}[H]
\vspace{-0.0cm}
    \centering
\begin{adjustbox}{width=0.5\textwidth}
\begin{tabular}{c}
\toprule
  Doc: Incorrect Focal Function \\
\midrule
\begin{minipage}[t]{0.45\textwidth}
\begin{minted}{text}
Write documentation for Vec
-----------------------------
\end{minted}
\begin{minted}[escapeinside=||]{cpp}
Vec Material::get_colour_at(double u, double v) const {
    if (m_texture.is_loaded())
        return m_texture.get_pixel(u, v);

    return m_colour;
}
\end{minted}
\end{minipage}\\
\midrule
GPT-3.5 Passing Response \\
\midrule
\begin{minipage}[t]{0.45\textwidth}

\begin{minted}[escapeinside=||]{cpp}
/**
 * @brief Retrieves the color at the specified texture coordinates.
 * 
 * @param u The u-coordinate of the texture.
 * @param v The v-coordinate of the texture.
 * @return The color at the specified texture coordinates.
 */
Vec Material::get_colour_at(double u, double v) const {
    if (m_texture.is_loaded())
        return m_texture.get_pixel(u, v);

    return m_colour;
}\end{minted}
\end{minipage} \\
\midrule
GPT-4 Failing Response \\
\midrule
\begin{minipage}[t]{0.45\textwidth}

\begin{minted}[escapeinside=||]{cpp}
/**
 * @class Vec
 * 
 * Represents a vector in 3D space. This class is typically 
 * used to represent points in 3D space or RGB color values.
 * 
 * Note: The actual definition of this class 
 * is not shown in this file.
 */
Vec Material::get_colour_at(double u, double v) const {
    if (m_texture.is_loaded())
        return m_texture.get_pixel(u, v);

    return m_colour;
}\end{minted}
\end{minipage} \\

\\
\bottomrule
\end{tabular}
\end{adjustbox}
\vspace{-0.0cm}
\caption{Example prompt and response for doc evaluation. GPT-3.5 passed, and GPT-4 failed because it added documentation for the Vec class, rather than the get colour function. This is caused by a failure we were able to highlight in the parsing and prompting of VSCode's pipeline: the model is erroneously asked to document Vec, not get colour. GPT-4 failed because it is better at following directions.}
\vspace{-0.0cm}
\label{fig:doc_case5401}
\end{figure}


With recommendations based on these results from the evaluation harness, we were able to significantly decrease the number of syntax and code logic errors in documentation generation in the LLM powered chat extension for VS Code. We were also able to fix the error in the parsing of functions in Java and C++ that resulted in the model receiving the incorrect keyword as the function name. Discoveries of such errors, as well as the respective improvements, are only possible with a robust and comprehensive evaluation system.

\subsubsection{Bug Fixing (fix)}
To understand the performance of the LLMs on the bug fixing evaluation, we delve deeper into the types of errors our static analyzers surface. Table \ref{tab:fix_errors_breakdown} shows a breakdown of some common static analyzer errors given to the models to solve for Typescript, as well as the GPT models' performance on them. We see that both models are often able to find the namespace of objects and fix type issues. However, as referenced previously, there are still many cases where the "has an 'any' type" error is not correctly resolved. Figure \ref{fig:case-1924} shows such an example. And, while Figure \ref{fig:case-1924} shows an example for GPT-4, Figure \ref{fig:case-1940} shows a similar phenomenon for GPT-3.5. The GPT-3.5 model hallucinates a type $Token$ for the $token$ variable. Tho tackle this issues in a LLM powered chat extension, we need to provide the models with additional context such as target variable types and namespaces so the LLM may correctly fix this issue, rather than incorrectly using existing types or hallucinating entirely new types.


\begin{table}[h]
    \centering
    \begin{tabular}{@{}llc@{}}
 \multicolumn{3}{c}{\textbf{Fix Errors in Typescript}}\\
\toprule
\textbf{Bug Type}&\textbf{Model}&\textbf{Bug Fixed}\\
\toprule
\multirow{2}{*}{Cannot find name/namespace}&  GPT-4 & 81\%\\&  GPT-3.5 & 70\%\\
\toprule
\multirow{2}{*}{Has an 'any' type}&  GPT-4 & 79\%\\&  GPT-3.5 & 64\%\\
\toprule
\multirow{2}{*}{Is not generic}&  GPT-4& 50\%\\&  GPT-3.5 & 54\%\\
\bottomrule
\\
\end{tabular}
\caption{Breakdown of common types of errors detected by the static analyzer for Typescript, as well as the relative performance of the GPT models on each type of error.}

\label{tab:fix_errors_breakdown}
\end{table}

\subsection{RQ3. Data Validation}
Although our dataset are pulled from real-world git repositories, this does not guarantee that our test cases accurately reflect how users interact with LLMs through IDEs. To confirm the validity of our dataset, we gather usage data that illustrates how hundreds of Microsoft developers use the docstring generation and bug fixing functionalities of our target LLM powered chat extension in VS Code. We then compare these instances to our test cases. \\

\begin{figure}[H]
\vspace{-0.0cm}
    \centering
\begin{adjustbox}{width=0.5\textwidth}
\begin{tabular}{c}
\toprule
  Fix Hallucination - GPT-3.5\\
\midrule
\begin{minipage}[t]{0.45\textwidth}
\begin{minted}{text}
You have been given the file contents of Parser.ts
The following code snippet within the file has a bug:
\end{minted}
\begin{minted}[escapeinside=||]{typescript}
private get categorizedTokens() {
    /**
     * @link http://stackoverflow.com/a/10939280/1149074
     */
    const parameterAssignmentTokens = _.takeWhile(
            this.tokens, token => token.value.includes("="));

    const commandWordToken = this.tokens[
                parameterAssignmentTokens.length];

    const beforeArgumentListTokensCount = 
                parameterAssignmentTokens.length + 1;
    const argumentListTokens = _.takeWhile(this.tokens.slice
                    (beforeArgumentListTokensCount), token => !(
        token instanceof Scanner.InputRedirectionSymbol ||
        token instanceof Scanner.OutputRedirectionSymbol ||
        token instanceof Scanner.AppendingOutputRedirectionSymbol
    ));
    ...

\end{minted}
\begin{minted}{text}
This is the line with the error:
\end{minted}
\begin{minted}[escapeinside=||]{typescript}
    const parameterAssignmentTokens = _.takeWhile(
            this.tokens, token => token.value.includes("="));
\end{minted}
\begin{minted}{text}
This is the problem with the line:
Parameter 'token' implicitly has an 'any' type.
Describe in a single sentence how you would solve the problem. 
Then, fix the error.
\end{minted}
\end{minipage}\\
\midrule
GPT-3.5 Failing Response \\
\midrule
\begin{minipage}[t]{0.45\textwidth}
\begin{minted}{text}
To fix the problem of the implicitly typed 'token' parameter, 
I would add type annotations to the parameters of the 
arrow functions in the code block
\end{minted}
\begin{minted}[escapeinside=||]{typescript}
private get categorizedTokens() {
    /**
     * @link http://stackoverflow.com/a/10939280/1149074
     */
    const parameterAssignmentTokens = _.takeWhile(
            this.tokens, (token: Token) => token.value.includes("="));

    const commandWordToken = this.tokens[
                parameterAssignmentTokens.length];

    const beforeArgumentListTokensCount = 
                parameterAssignmentTokens.length + 1;
    const argumentListTokens = _.takeWhile(this.tokens.slice
                    (beforeArgumentListTokensCount), (token: Token) => !(
        token instanceof Scanner.InputRedirectionSymbol ||
        token instanceof Scanner.OutputRedirectionSymbol ||
        token instanceof Scanner.AppendingOutputRedirectionSymbol
    ));
    ...
\end{minted}
\end{minipage} \\
\bottomrule
\end{tabular}
\end{adjustbox}
\vspace{-0.0cm}
\caption{Example failing prompt and response from GPT-3.5. The model hallucinates a type (Token) that matches the name of the variable.}
\vspace{-0.0cm}
\label{fig:case-1940}
\end{figure}

\begin{figure}[h]
  \includegraphics[width=0.45\textwidth]{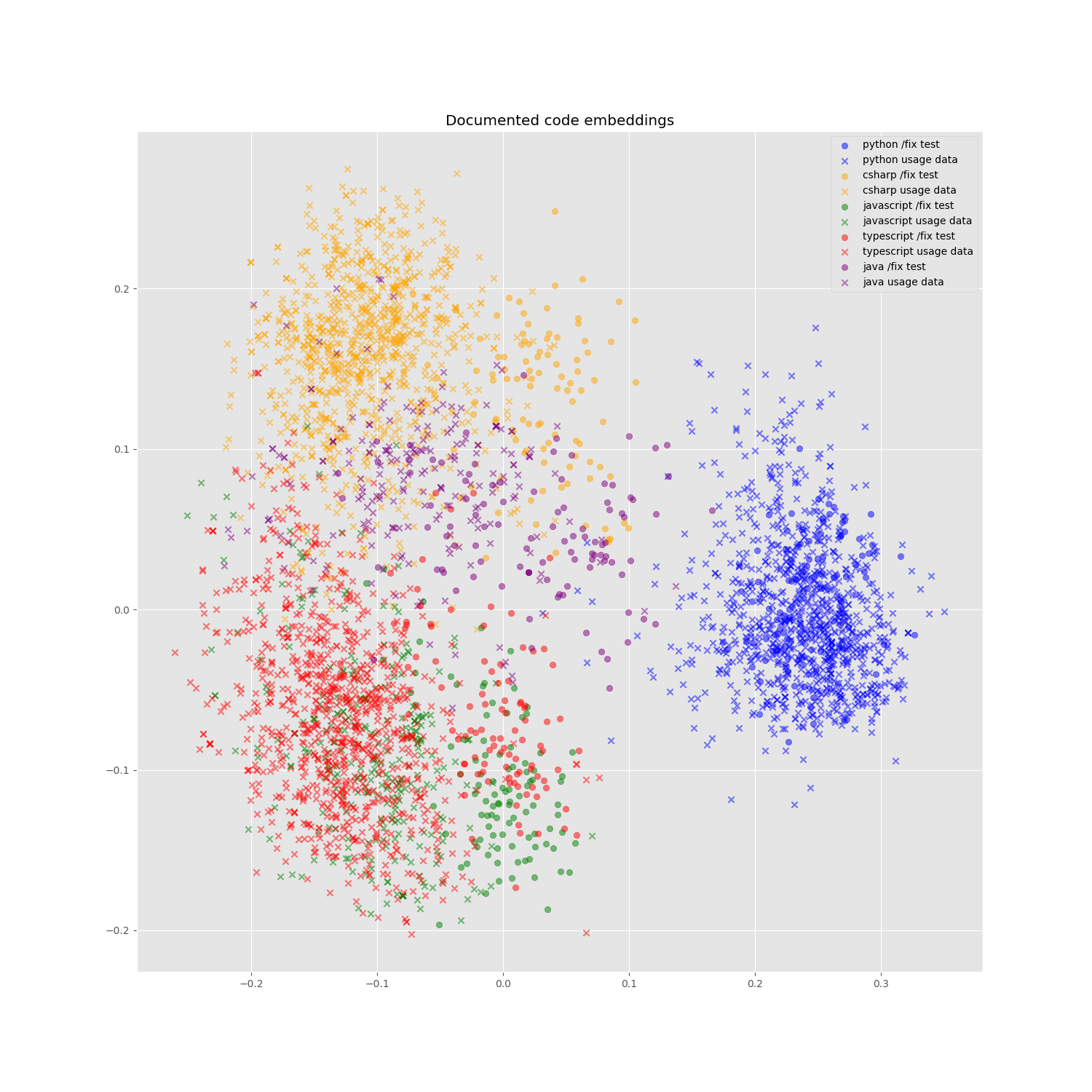}
\caption{Comparing our dataset for Documentation generation evaluation with real-world usage across languages.}
\label{fig:telemetry-doc}
\end{figure}

\begin{figure}[h]
  \includegraphics[width=0.45\textwidth]{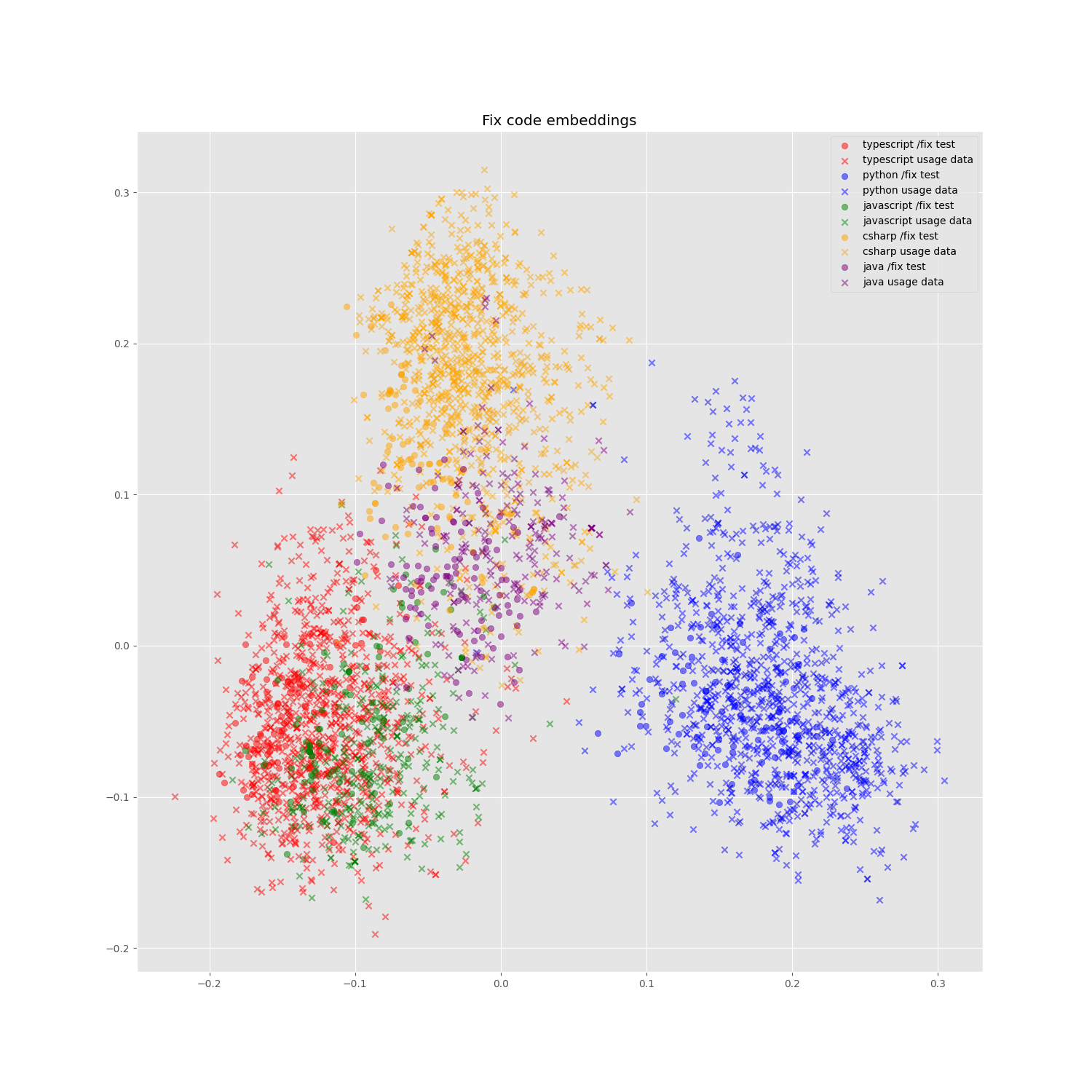}
\caption{Comparing our dataset for Bug Fixing evaluation with real-world usage across languages.}
\label{fig:telemetry-fix}
\end{figure}

For documentation generation, we use OpenAI's ada embedding model \cite{ada} to embed the documented code snippets and compare the snippets from our dataset with the one's gathered from Microsoft developers usage data. Similarly, for the bug fixing telemetry, we embed the code snippets that contain the bug. We use PCA dimensionality reduction to plot the data in two dimensions. PCA dimensionality reduction is optimized to find a plane that maximizes the distance between points and outliers. Figures \ref{fig:telemetry-doc} and \ref{fig:telemetry-fix} show the results of this comparison. We see that each language forms a cluster, and the real usage and our data exist within a similar space for each language cluster, for both documentation generation and bug fixing.

We do not aim to match the test cases in our dataset to the real usage point for point. Rather, we are determining whether or not our test cases are outliers in the space of the real usage. If they are not outliers, we can infer that our dataset is in line with the real-world usage of the chat extension. This analysis suggests that our dataset for both the documentation generation and bug fixing evaluation is in line with real world usage.

\section{Conclusion and Future Work}
With the growing use of LLMs to aid developers in complex engineering tasks comes the need for more robust evaluations of LLM-generated code. Especially as more companies and products seek to integrate LLMs into their workflows, existing evaluation metrics are not sufficient to confirm the quality and correctness of machine-generated code. In this paper, we propose a solution to this problem via the Copilot Evaluation harness. We define five key evaluation metrics for the code generation problem space: method generation, test generation, docstring generation, bug fixing and workspace understanding. We detail the methodology required to collect test cases and evaluation results for each of those five metrics. We also provide preliminary results for two of the five metrics across myriad programming languages.

Our goal in creating the evaluation harness is to validate the quality of LLM-generated code. Although we have seen immense advancements in the code generation ML space, we seek to highlight how much oversight and engineering effort is required to reliably and optimally integrate LLMs into a code workflow. We aim to provide developers a comprehensive evaluation suite, with which they can optimize the integration of LLMs into their coding workflows. With the Copilot Evaluation harness, programmers can more systematically and robustly evaluate the impact of parameters such as prompt wordings, changes in the order of information provided, changes in the context provided to the model, and more.

Moreover, the Copilot Evaluation harness can be used for cost optimizations by revealing that a more budget-friendly LLM model (e.g. CodeLLama) might exhibit satisfactory performance in tasks like documentation. This insight enables developers to intelligently balance resources by allocating tasks to the cost-effective LLM when its performance is deemed sufficient. Simultaneously, more complex tasks can be shifted to more powerful LLMs to ensure optimal outcomes.

We publish this paper as a living documentation of our progress. Future work on this project involves reporting results for the remaining three evaluation metrics and open-sourcing our data and evaluation code.

\bibliographystyle{acm}
\bibliography{main}



\end{document}